\documentclass[twocolumn,prb,showpacs,superscriptaddress]{revtex4}  

\usepackage{graphicx}
\usepackage{amssymb}

\begin{document}
\def\k {{\bf k}}
\def\n {{\hat{\bf n}}}
\def\r {{\bf r}}
\def\u {{\bf u}}
\def\e {{\bf e}}

\def\D {{\bf D}}
\def\G {{\bf G}}
\def\H {{\bf H}}
\def\A {{\bf A}}

\title{Nature of vibrational eigenmodes in topologically disordered
solids}

\author{S. I. Simdyankin} 

\affiliation{D\'epartement de physique, Universit\'e de Montr\'eal,
C.P. 6128, succ. Centre-ville, Montr\'eal (Qu\'ebec) H3C 3J7, Canada}

\affiliation{Department of Numerical Analysis and Computer Science,
Royal Institute of Technology, SE-100 44 Stockholm, Sweden}

\author{S. N. Taraskin} 
\affiliation{Department of Chemistry, University of Cambridge,
             Lensfield Road, Cambridge CB2 1EW, United Kingdom}

\author{M. Elenius}
\affiliation{Department of Numerical Analysis and Computer Science, 
             Royal Institute of Technology, SE-100 44 Stockholm, Sweden}

\author{S. R. Elliott} 
\affiliation{Department of Chemistry, University of Cambridge,
             Lensfield Road, Cambridge CB2 1EW, United Kingdom}

\author{M. Dzugutov}
\affiliation{Department of Numerical Analysis and Computer Science, 
             Royal Institute of Technology, SE-100 44 Stockholm, Sweden}

\date{\today}

\begin{abstract}
We use a local projectional analysis method to investigate the
effect of topological disorder on the vibrational dynamics in a model
glass simulated by molecular dynamics.
Evidence is presented that the vibrational eigenmodes in the glass are
generically related to the corresponding eigenmodes of its crystalline
counterpart via disorder-induced level-repelling and hybridization
effects.
It is argued that the effect of topological disorder in the
glass on the dynamical matrix can be simulated by introducing
positional disorder in a crystalline counterpart.
\end{abstract}

\pacs{63.50.+x, 
      63.20.Dj, 
      61.43.-j  
     }

\maketitle

%
%
\section{Introduction}
\label{s1}
Vibrational properties of disordered materials is one of the current
lively topics of modern condensed-matter physics \cite{Elliott_PAM}.
Features of the disordered vibrational spectrum, such as the boson
peak, the Ioffe-Regel crossover and vibrational localization are being 
investigated (see e.g.
\cite{Schirmacher_98,Grigera_01,Martin_Mayor_01:JCP,Taraskin_00:PRB_IR1,Taraskin_00:PRB_IR2,Taraskin_01:PRL}).

Disorder in condensed matter can essentially be classified into two
basic types \cite{Ziman_MD}:
(i) lattice disorder in crystals, e.g. in substitutional alloys, and
(ii) topological disorder in, e.g., glasses.
Crystalline structures with lattice disorder can be studied
analytically and considerable progress has been achieved in
understanding their vibrational behaviour
\cite{Schirmacher_98,Martin_Mayor_00,Taraskin_01:PRL}. 
Structures with topological disorder have mainly been investigated
numerically but in some aspects analytically 
\cite{Wu_92,Wan_94,Biroli_99,Monasson_99,Martin_Mayor_00,Barrat_00,Grigera_01,Grigera_01:BP}. 
Different possible ways of describing the origin and nature of
vibrational modes, particularly in the low-frequency range, 
 in such disordered systems are still under debate.

The vibrational properties of crystals with lattice disorder can be
successfully treated mainly because of the existence of a well-defined
reference structure, viz. the same crystalline structure but without
disorder.
The choice of a reference structure for a topologically disordered
material is less obvious.
In some cases, such a choice can be based on the existence of similar
local order in both the topologically disordered structure and its
crystalline prototype.
For example, $\alpha$-cristobalite is found to be a good crystalline
counterpart for vitreous silica \cite{Taraskin_97:PRB2,Ding_98}.
The existence of similar structural elements 
in both crystalline and glassy systems 
is expected to lead to
similar vibrational dynamics.
Similarities in the vibrational dynamics within these structural units
can be revealed by comparing vibrational eigenmodes of the
topologically disordered structure with those of its crystalline
counterpart.
Such a comparison can elucidate the nature of the disordered
vibrational eigenmodes and their possible generic connection to
crystalline modes.
Basically, it can be expected that vibrational eigenmodes in
topologically disordered structures have approximately the same nature 
as in crystals with lattice disorder
\cite{Taraskin_99:PRB,Taraskin_01:PRL}.
Namely, they can be regarded as strongly hybridized crystalline
eigenmodes, shifted in frequency due to disorder-induced
level-repelling effects.
In this paper, we present arguments, based on a numerical analysis of
a representative topologically disordered structure, that this
conjecture holds true, at least, for the structure considered.

The existence of a crystalline counterpart allows a comparison to be
made between a glass and its reference structure and conclusions to be
drawn about the effect of topological disorder on the spectrum of
vibrational excitations.
Moreover, it is possible to compare positional, e.g. 
quenched thermally-induced, 
disorder in a crystal with topological disorder in the corresponding
glass in terms of their respective influence on the vibrational
properties.
In a computer simulation, positional disorder can be introduced into a
crystalline structure by heating it in the course of a 
molecular-dynamics run up to a certain temperature below the melting point.
Then the dynamical matrix of an instantaneous configuration
corresponding to this temperature can be calculated and analyzed.
Thus it is possible to mimic the main features of the vibrational
dynamics of topologically disordered systems by the vibrational
dynamics of their positionally disordered crystalline counterparts.
In the following, we demonstrate that this can be done, at least for
some representative structures.

The rest of the paper is arranged as following. 
The local projectional analysis method is developed and described 
in Sec.~\ref{s2}. 
The model is described in Sec.~\ref{s22}. 
Results and conclusions are given in Sec.~\ref{s3} 
and  ~\ref{s4}, respectively. 

\section{Local projectional analysis} 
\label{s2}

A comparison of the vibrational dynamics in a glass and its
counterpart crystalline phase can be based on a local projectional
analysis.
Of course, the equilibrium atomic arrangements in the glass and the
crystal are essentially different, and thus we cannot expand the
eigenmodes in a disordered system in terms of the crystalline
eigenmodes, as we can do for disordered lattices
\cite{Taraskin_01:PRL}.
Nevertheless, what we can do in such a situation is to compare the
local atomic motions in topologically similar constituent structural
units 
in different frequency ranges in the glass and the crystal.

The local projectional analysis uses the information about atomic
displacements contained in the vibrational eigenmodes $\{\e_{\xi,i}\}$
with eigenfrequencies $\{\omega_{\xi}\}$.
Here the index $i$ enumerates the atoms and the index $\xi$ labels the
eigenmodes.
Let us introduce the normalized local displacement vector
$\{\u^{(l)}_{\xi,j}\}$ for a structural element $l$: 
$\u^{(l)}_{\xi,j} = \e_{\xi,j}/[p_{l}(\omega_{\xi})]^{1/2}$, 
where $j$ enumerates the atoms within the structural
unit, and $p_{l}(\omega_{\xi}) = \sum_{j} |\e_{\xi,j}|^2$.
In the case of a crystal, an eigenmode is 
identified by the wavevector $\k$ and the dispersion branch $\beta$.
For a crystalline structural unit $l$, a disordered structural unit
$m$, a crystalline eigenmode $\k\beta$ and a disordered eigenmode
$\mu$,  we can define the squared scalar product:
\begin{equation}
A_{l,m}(\omega_{\k\beta},\omega_{\mu}) = |\u^{(l)}_{\k\beta}
\u^{(m)}_{\mu}|^2~,  
\label{e0}
\end{equation}
where $\u^{(l)}_{\k\beta}$, $\u^{(l)}_{\mu}$ have $3N_u$ components,
$N_u$ being the number of atoms within the structural unit.  
This quantity depends on the mutual orientation of the crystalline and
disordered structural units.
We are interested in finding the crystalline eigenmode 
which most resembles a given  disordered mode  
and, therefore, we choose the maximum 
value (among all possible mutual orientations preserving 
the symmetry) of the squared scalar product, 
\begin{equation}
{\tilde A}_{l,m}(\omega_{\k\beta},\omega_{\mu})=
{\rm max}\{A_{l,m}(\omega_{\k\beta},\omega_{\mu})\}~. 
\label{e00}
\end{equation}
The next step is to average 
${\tilde A}_{l,m}(\omega_{\k\beta},\omega_{\mu})$ 
over all possible disordered and crystalline structural units, 
and thus calculate the averaged squared scalar product, 
$ {\overline A}(\omega_{\k\beta},\omega_{\mu})$, 
\begin{equation}
{\overline A}(\omega_{\k\beta},\omega_{\mu}) 
= \frac{1}{P}\sum_{l,m} p_l(\omega_{\k\beta}) 
p_m(\omega_{\mu}) 
{\tilde A}_{l,m}(\omega_{\k\beta},\omega_{\mu}) 
\ , 
\label{e1}
\end{equation}
where the local normalization factors, $p_{l(m)}$, 
take into account the weight of the corresponding 
structural unit in the entire mode and 
$P\equiv \sum_{l,m} p_l(\omega_{\k\beta}) 
p_m(\omega_{\mu})$. 
The averaged scalar products ${\overline
A}(\omega_{\k\beta},\omega_{\mu})$ can be reduced to a function of two
continuous variables ${\overline A}(\omega_{\rm cryst},\omega_{\rm
dis})$ by an interpolation.
In this case, ${\overline A}(\omega_{\rm cryst},\omega_{\rm dis})$
quantifies the degree of local similarity between one disordered mode
with a frequency about $\omega_{\rm dis}$ and a crystalline mode with
a frequency about $\omega_{\rm cryst}$.
One way to obtain ${\overline A}(\omega_{\rm cryst},\omega_{\rm dis})$
from ${\overline A}(\omega_{\k\beta},\omega_{\mu})$, which was used in
the present work, is to compile a histogram for a discrete set of
equidistant values of $\omega_{\rm cryst}$ for an arbitrary set of
$\omega_{\mu} \equiv \omega_{\rm dis}$, 
i.e. for each $\omega_{\rm dis}$, 
\begin{equation}
{\overline A}(\omega_{\rm cryst},\omega_{\rm dis}) =
\frac{1}{N(\omega_{\rm cryst})} 
 \sum_{\k'\beta'}{\overline
A}(\omega_{\k'\beta'},\omega_{\mu})~. 
\label{e11}
\end{equation}
Here $\k'\beta'$ assume values for which $\omega_{\rm cryst} \le
\omega_{\k'\beta'} < \omega_{\rm cryst}+d\omega$, where $d\omega$ is
the width of a histogram bin, and $N(\omega_{\rm cryst})$ is the number of
crystalline states in this spectral interval of width $d\omega$.

\section{Structural model}
\label{s22}

As a representative example of a topologically disordered model
structure, we consider a single-component glass with predominantly
icosahedral order (the IC glass) constructed by means of 
molecular-dynamics simulation with the use of 
a pair-wise interatomic potential
\cite{Dzugutov_92} (all quantities used in this paper are expressed in
 Lennard-Jones reduced units \cite{Allen_CSL}, see
Refs.~\onlinecite{Simdyankin_01:PRB,Simdyankin_00} for more detail).
The $3N\times 3N$-dynamical matrix (the number of particles is
$N=16000$) for a glassy minimum-energy configuration has been
calculated and diagonalized, yielding all vibrational eigenvectors,
$\{\e_{\mu,i}\}$, and eigenfrequencies, $\{\omega_{\mu}\}$
for this system.
A good crystalline counterpart for the IC glass is the $\sigma$ phase,
a Frank-Kasper crystal \cite{Simdyankin_00,Harrop_01}.

For the local projectional analysis of the type described above, we
have chosen two 
$Z14$
Frank-Kasper polyhedra (point group $D_{6h}$), 
interpenetrating along the six-fold symmetry axes.
This structural unit ($Z14_2$),  comprised of 22 atoms,  represents a
short segment of the -72$^{\circ}$ disclination line
\cite{Mosseri_GF,Simdyankin_01:PRB}.
Within the unit cell of the $\sigma$ phase, there are 16 interior
atoms (centers of $Z14$ polyhedra) of eight partially overlapping
$Z14_2$ units.
In the 16000-atom sample of the simulated IC glass, 236 such units
(some of them partially overlapping) covering 20\% of all atoms have
been identified.
These atoms, together with their nearest neighbors,  represent 52\% of
all atoms.

\section{Results}
\label{s3}

\begin{figure} 
\centerline{\includegraphics[width=7.cm]{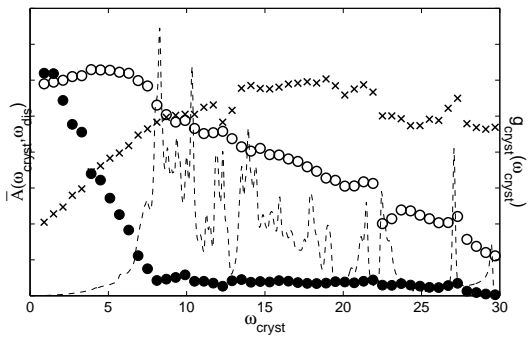}}
\caption{ Averaged squared scalar product of the atomic
displacements of disordered and crystalline local units renormalized
to the maximum value as a function of the crystalline frequency
$\omega_{\rm cryst}$ for three values of the disordered mode frequency
$\omega_{\rm dis}$: {\footnotesize $\bullet$}; 1.12, $\circ$; 6.84;
$\times$, 16.81. Dashed line: vibrational density of states $g_{\rm
cryst}(\omega_{\rm cryst})$ for the $\sigma$ phase.  
}
\label{f1}
\end{figure}

\begin{figure} 
\centerline{\includegraphics[width=7.5cm]{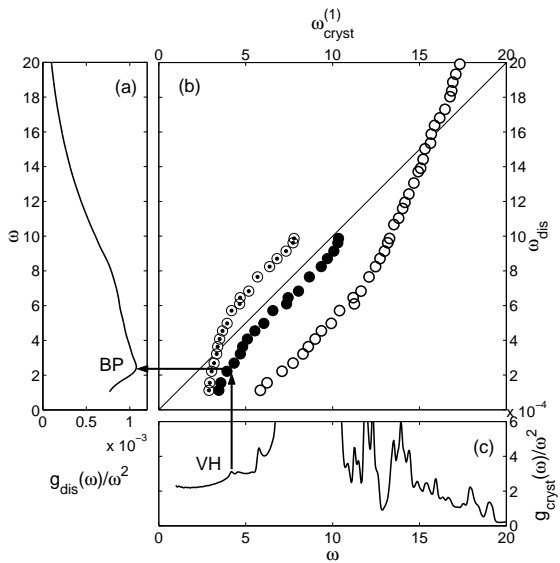}}
\caption{ (a) Reduced vibrational density of states,
$g_{\text{dis}}(\omega)/\omega^2$, for the IC glass showing the boson
peak (BP).  (b) Disorder-induced level-repelling effect: the
disordered mode frequency $\omega_{\rm dis}$ versus the characteristic
crystalline frequency $\omega^{(1)}_{\rm cryst}$ evaluated using two
methods: $\circ$, direct; {\footnotesize $\bullet$},
background-subtracted with $\omega_0 \simeq 20$; {\tiny $\odot$},
background-subtracted with $\omega_0 \simeq 10$, see
Ref.~\onlinecite{background}.
The arrows show the connection between the
position of the lowest van Hove singularity (VH) and the boson peak
(BP) via the local projectional analysis results.  (c) Reduced
vibrational density of states for the $\sigma$ phase.}
\label{f11}
\end{figure}

\subsection{Level-repelling effect}
\label{s3b}

Fig.~\ref{f1} shows the calculated ${\overline A}(\omega_{\rm
cryst},\omega_{\rm dis})$ for disordered eigenmodes from the {low-,}
middle- and high-frequency parts of the spectrum (a set of crystalline
eigenmodes uniformly covering the first Brillouin zone with $10^3$
points in $\k$-space was used).
As seen from this figure, ${\overline A}(\omega_{\rm
cryst},\omega_{\rm dis})$, as a function of $\omega_{\rm cryst}$,
represents a broad distribution centered around $\omega_{\rm cryst} =
\omega_{\rm cryst}^{\rm max}$, the frequency of the crystalline
spectrum for which the cystalline modes have the greatest overlap with the
modes of the disordered system.
This suggests that a disordered mode with a frequency about
$\omega_{\rm dis}$ is generically related to crystalline modes about
$\omega_{\rm cryst}^{\rm max}$; in other words, the vibrational motion
of the $Z14_2$ structural units in the disordered eigenmode is similar
to that of the same structural units for the indicated eigenmodes in
the $\sigma$ phase.
Since ${\overline A}(\omega_{\rm cryst},\omega_{\rm dis})$ as a
function of $\omega_{\rm cryst}$ is broad and asymmetric, particularly at high 
frequencies, we estimate
the characteristic frequency of the crystalline modes corresponding to
a disordered mode associated with $\omega_{\rm dis}$ as the first
moment of this function:
\begin{equation}
\omega_{\rm cryst}^{(1)}(\omega_{\rm dis}) = 
\int \omega {\overline A}(\omega,\omega_{\rm dis}) d\omega / 
\int {\overline A}(\omega,\omega_{\rm dis}) d\omega~. 
\label{e111}
\end{equation}
In Fig.~\ref{f11}(b), $\omega_{\rm cryst}^{(1)}$, calculated both by
the direct method and by subtracting a background from ${\overline
A}(\omega_{\rm cryst},\omega_{\rm dis})$ \cite{background}, is plotted
versus $\omega_{\rm dis}$.

From Fig.~\ref{f1}, it is evident that the peak-shaped function,
${\overline A}(\omega_{\text{cryst}}, \omega_{\text{dis}})$, becomes
increasingly broader with increasing frequency,
$\omega_{\text{cryst}}$, meaning that there is an increasingly weak
correlation in character between modes in the disordered system and
specific similar modes in the crystal.  
Such a large broadening of the
peaks at high frequencies has also been found in lattice models
\cite{Taraskin_01:PRL} subject to large force-constant disorder and,
in this sense, it is not surprising and reflects the large degree of
disorder in the IC-glass.
A similar effect is known for vibrational plane waves (analog of
crystalline eigenmodes) propagating in glass.
The distribution of the weights of different disordered eigenmodes
contributing to a propagating plane wave becomes increasingly broad
with increasing plane-wave frequency (the strong-scattering regime)
\cite{Taraskin_00:PRB_IR1,Taraskin_00:PRB_IR2}.
The shape of this distribution can even approach the shape of the
entire frequency spectrum of a glass.
However, at low frequencies, and specifically in the boson-peak
region, $\omega_{\text{BP}} \simeq 2.5$, the peak width of ${\overline
A}(\omega_{\text{cryst}}, \omega_{\rm dis})$ is rather narrow,
indicating that there is quite a strong correlation between the two
types of modes there (this is analogous to the weak-scattering regime
for plane waves \cite{Taraskin_00:PRB_IR1,Taraskin_00:PRB_IR2}).

The deviation of the curve $\omega_{\text{cryst}}^{(1)}(\omega_{\rm
dis})$ from the straight-line bisector,
$\omega_{\text{cryst}}^{(1)}=\omega_{\text{dis}}$, reflects the
level-repelling effect.
It is clearly seen that the disordered modes from the low-frequency
region are pushed down in frequency as compared to the related
crystalline frequency, while the disordered modes from the
high-frequency part of the spectrum are shifted upwards with respect
to $\omega^{(1)}_{\rm cryst}$.

Why do we describe this phenomenon as a level-repelling effect?  One
can argue alternatively that disordered sytems are ``softer'' than
their crystalline counterparts, and this results in a smaller sound
velocity and in  excess low-frequency eigenmodes which have nothing
to do with the level-repelling effect.
In the high-frequency range, the extra modes could be due to
vibrations in local structural blocks characterized by relatively
large (due to statistical fluctuations) spring constants which are
also not related to the level-repelling effect.  
These arguments are not based on particular asumptions about the
topology of the disordered system and therefore should be valid for
disordered lattices as well.
However, we have demonstrated the relevance of the level-repelling
effects in describing vibrational spectra of disordered lattices with
force-constant disorder \cite{Taraskin_01:PRL}.  
This has been done within the mean-field approximation (homogeneous
disorder), ignoring the role of statistical fluctuations in the
force-constant distributions.
We have explained the appearance of 
the low-frequency 
excess modes (with the VDOS $\propto \omega^2)$ 
in terms of the well-known effect \cite{Ehrenreich_76} 
of the shift of the low-frequency part of the spectrum 
of quasi-particles to lower 
frequencies due to level-repelling effects. 
The mean-field approach is very accurate in the 
low-frequency regime. 
This demonstrates the  relative unimportance of  local 
soft regions (soft configurations) characterized by  
relatively small force constants which 
are due to statistical fluctuations. 
These regions, of course, exist but the low-frequency 
eigenmodes are not localized at them due to the 
delocalized  properties of the vibrational 
states  spectrum around zero frequency 
\cite{John_83,Taraskin_99:PRB}. 
We have also examined and 
confirmed this by means of multifractal 
analysis (to be published elsewhere). 
This is in contrast to the high-frequency part of the VDOS 
where the mean-field approach describes well only 
the main features and tendencies in the spectrum, such as the  
shift to  higher frequencies (again due to level-repelling effects) 
but not the band tails 
containing 
localized states where  statistical fluctuations 
of the force constants are very important. 

\begin{figure} 
\centerline{\includegraphics[width=7cm]{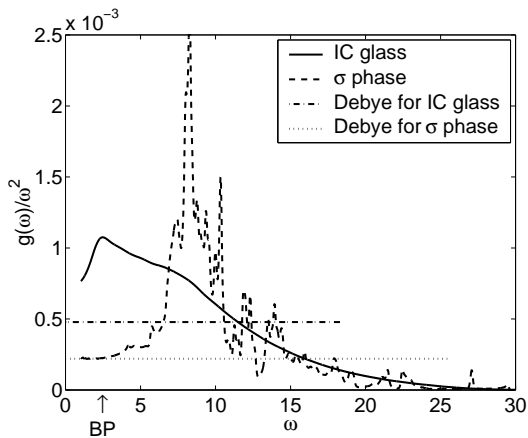}}
\caption{The reduced vibrational densities of states $g_{\rm
dis}(\omega)/\omega^2$ and $g_{\rm cryst}(\omega)/\omega^2$, for the
IC glass and the $\sigma$ phase, respectively.  The position of the
boson peak (BP) in the VDOS of the IC glass is marked by the arrow.}
\label{f1.1}
\end{figure} 

Thus, bearing in mind the importance of  
level-repelling effects in determining   the 
spectrum of disordered lattices, we 
conjecture  that a similar effect can also be important 
in topologically disordered  structures. 
By this we mean  level-repelling effects relative to  the 
spectra of corresponding crystalline counterparts of 
disordered  sytems which are caused by 
positional and topological disorder. 
The crystalline counterparts are characterized by the 
same local order as in the corresponding disordered 
structures. 
In topologically disordered systems, 
the atoms in local structural units are 
displaced from their ideal crystalline positions 
 and the values of the force-constants are distributed,  
giving a situation 
resembling  lattice models with force-constant disorder. 
However, 
the force constants of bonds emanating from a given central atom 
can be correlated with each other (see also Ref.~\onlinecite{Mezard_99}).  
This situation is in contrast to lattice models with uncorrelated 
force constants. 
The strength and role of these correlations in the vibrational spectrum 
are not presently known. 
Another distinctive feature of glassy systems is the presence of 
topological disorder, the role of which in  vibrational 
spectra is not yet established. 
In our current analysis, we try to find out to what extent  
the ''lattice ideology'' (including the level-repelling 
effect) is applicable to glassy models.  

\subsection{The boson peak}
\label{s3_bb}

\begin{figure} 
\centerline{\includegraphics[width=8.5cm]{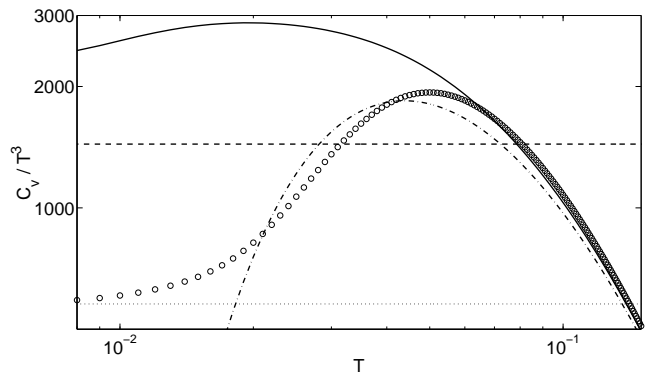}}
\caption{ Calculated values of the temperature dependence of the
vibrational contribution to the heat capacity for the IC glass (solid
line) and the $\sigma$-phase crystalline approximant (circles),
plotted as $C_{\text{v}}/T^3$.  
The curve for the IC glass calculated by omitting all the
states with $\omega < 4$ (i.e. in the boson-peak region) is also shown
by the dot-dashed curve.  The calculated Debye values for
$C_{\text{v}}/T^3$ for the IC glass (dashed line) and $\sigma$-phase
(dotted line) are also given.  }
\label{Cv}
\end{figure}

Fig.~\ref{f11}(b) can be used to establish the nature of disordered
modes in different frequency ranges.
The low-frequency regime is of particular interest for vibrations in 
glasses, because of the so-called boson peak (BP) which occurs in the
reduced VDOS, $g_{\rm dis}(\omega)/\omega^2$, [see Figs.~\ref{f11}(a) 
and \ref{f1.1}] 
and in the temperature dependence of the (vibrational) heat 
capacity $C_{\text{v}}$, normalized by the Debye $T^3$ dependence 
[the solid line in Fig.~\ref{Cv}].

The vibrational contribution to the heat capacity 
per particle is given by
\begin{equation}
C_{\text{v}}(T) = \frac{1}{N}\sum_{j=1}^{3N} \left(
\frac{\hbar\omega_j}{T} \right)^2
\frac{\exp(\hbar\omega_j/T)}{\left[\exp(\hbar\omega_j/T) - 1
\right]^2} .
\end{equation}
Since all quantities in this paper are expressed in  Lennard-Jones
reduced units, we use the value of $\hbar\simeq0.02959$ referred to
argon.
The Debye approximation of the above equation is given by
\begin{equation}
C_{\text{v}}^D(T) = \frac{12\pi^4}{5} 
\left( \frac{T}{\theta_D} \right)^3 ,
\end{equation}
where $\theta_D = \hbar\omega_D = 2\pi\hbar (9N/4\pi V)^{1/3} c_D$ is
the Debye temperature \cite{Maradudin_TLDHA}.
The Debye velocity of sound $c_D$ is defined as $c_{\text{D}} =
(1/c_l^3 + 2/c_t^3)^{-1/3}$, where $c_l$ and $c_t$ are the
longitudinal and transverse sound velocities respectively, which can
be estimated from the slopes of the acoustic dispersion branches
\cite{Simdyankin_01:PRB}.
The Debye velocities for the IC glass and the $\sigma$ phase 
are 
$c_{\text{D}}^{\text{dis}} \simeq 3.43$ 
and
$c_{\text{D}}^{\text{cryst}} \simeq 4.74$,  
respectively.

In Fig.~\ref{Cv}, the peak in the curve of $C_{\text{v}}/T^3$ versus
$T$ is due to states in the boson-peak range of frequencies in the
vibrational spectrum.
Indeed, if we cut the low-frequency part 
of the spectrum, including the boson peak, then 
the peak 
in $C_{\text{v}}/T^3$ is significantly suppressed (see the 
dash-dotted line in  Fig.~\ref{Cv}). 
The heat capacity of the IC-glass can be easily compared with that of the  
$\sigma$-phase crystal. 
It can be seen from Fig.~\ref{Cv} that the plot 
of $C_{\text{v}}/T^3$  for the $\sigma$-phase crystal 
[the circles in Fig.~\ref{Cv}] 
tends to the Debye limit (the dotted line) 
at low temperatures. 
The increase above this level at higher $T$ is 
due to the influence of van Hove singularities 
in the crystalline VDOS. 
This peak for the $\sigma$-phase crystal closely resembles 
the peak for the IC-glass, 
when the modes in the boson-peak region 
of the VDOS are omitted from the calculations. 
However, the peak in the total curve of $C_{\text{v}}/T^3$ for the IC
glass is significantly shifted to lower frequencies, and also
approaches a higher value of the Debye limit (the dashed line in
Fig.~\ref{Cv}) because of a lower value of the sound velocity.

Following the arrows in Figs.~\ref{f11}(a)-(c), it appears that the
vibrational states in the boson-peak region mainly 
correspond to crystalline states in the 
vicinity of  the
lowest van Hove singularity in the $\sigma$ phase.
Although this method of analysis is unavoidably approximate, it does
seem to confirm the physical picture for the origin of the boson peak
previously found in the f.c.c. force-constant disordered lattice
\cite{Taraskin_01:PRL} and in models of disordered silicon
\cite{Finkemeier_01}.

Assuming such a scenario for the boson-peak origin, we can compare its
position, $\omega_{\text{BP}} \simeq 2.5 $, with the position of the lowest van Hove
singularity, $\omega_{\rm VH} \simeq 4$, and thus conclude that $
\omega_{\rm VH} - \omega_{\rm BP} \sim \omega_{\rm VH}$, which is an
indication of strong disorder. 
In this model,  
level repelling of low-frequency optic modes to even 
lower frequencies  significantly contributes 
to the excess mode density in the 
boson-peak region (see below). 

\begin{figure} 
\centerline{\includegraphics[width=7.5cm]{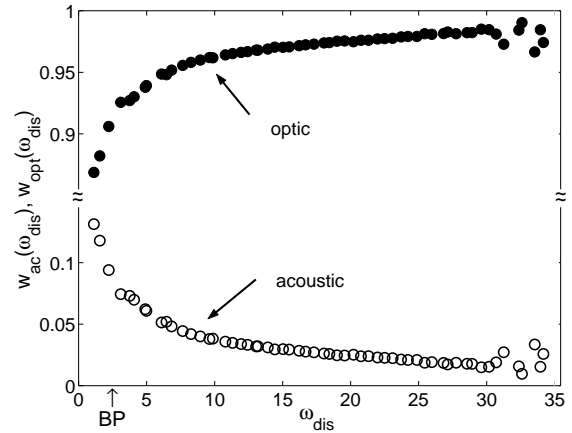}}
\caption{Hybridization parameters for acoustic ($\circ$) and optic
({\footnotesize $\bullet$}) branches versus the 
disordered mode frequency $\omega_{\rm dis}$. 
The boson-peak (BP) frequency is marked by arrow. 
} 
\label{f2}
\end{figure} 

\subsection{Branch-hybridization effects}
\label{s3c}

The level-repelling effect in glass is accompanied by another general effect,
namely strong disorder-induced hybridization of crystalline eigenmodes.
In this effect, many crystalline eigenmodes contribute to a particular
disordered eigenmode (cf. the broad distributions of ${\overline
A}(\omega_{\rm cryst},\omega_{\rm dis})$ in Fig.~\ref{f1}).
Disorder-induced hybridization also involves hybridization between  
different crystalline vibrational branches.
A quantitative characteristic of such hybridization can be defined
via the branch hybridization parameter, $w_\beta(\omega_{\mu})$:
\begin{equation}
w_{\beta}(\omega_{\mu})= 
\sum_\k {\overline A}(\omega_{\k\beta},\omega_{\mu}) 
/ \sum_{\k\beta} {\overline A}
(\omega_{\k\beta},\omega_{\mu})
\ .   
\label{e2} 
\end{equation}
Knowing $w_\beta(\omega_{\mu})$, we can 
say which crystalline branch mainly
contributes to a particular disordered eigenmode.
In particular, this is of interest for the low-frequency regime, where
only the acoustic branches exist in the crystal, but where we could also 
expect an admixture of 
low-lying optic branches in a glass due to hybridization
effects.
The results of a calculation of the hybridization parameter for
acoustic, $w_{\rm ac}=\sum_{\beta=1}^3 w_{\beta}$, and optic, $w_{\rm
opt}=\sum_{\beta=4}^{3N_{\rm u.c.}} w_{\beta}$, branches are presented
in Fig.~\ref{f2} (with $N_{\rm u.c.}$ being the number of atoms in the
crystalline unit cell; $N_{\rm u.c.}=30$ for the $\sigma$ phase). 
Here we have averaged the hybridization parameter 
$w_\beta$ over all acoustic and all optic branches 
in order to demonstrate the relative role of  
these types (acoustic or optic) vibrations in disordered modes. 
As seen from Fig.~\ref{f2}, even for the lowest-frequency disordered
modes, the contribution of optic crystalline modes is dominant
[$w_{\text{opt}}(\omega_{\text{dis}}\simeq 1) \simeq 0.8$] which
indicates very strong hybridization effects with the 
dominant number of optic modes.
Precisely because of such strong hybridization, it is not possible to
distinguish between the contributions of acoustic branches of
different polarizations (transverse or longitudinal).
The branches of different polarization are distinct for small
wavevectors only, but this part of the dispersion is suppressed by the
dominant contribution from all other wave vectors for which the
polarization is not well defined.
It is also seen in Fig.~\ref{f2} that the contribution of the acoustic
branches is maximum at low frequencies and then monotonically decays
with increasing frequency (whereas the opposite behavior is,
obviously, found for optic modes), thus indicating an enhancement of
the hybridization with optic branches.
Such behaviour is qualitatively different from the plateau behaviour
found at low frequencies in the f.c.c. disordered lattice
\cite{Taraskin_01:PRL}, where only acoustic modes exist, and is due to
the existence of the extensive optic spectrum in the crystalline
counterpart of the IC glass.

\begin{figure} 
\centerline{\includegraphics[width=7cm]{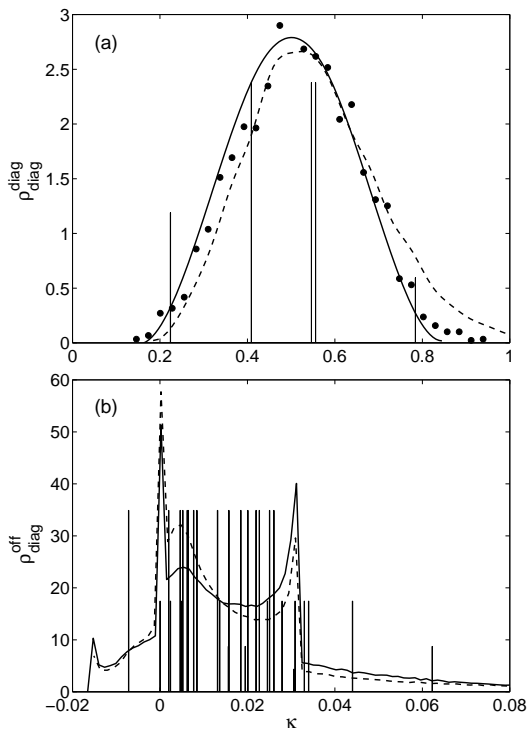}}
\caption{Distributions of the dynamical matrix elements, scaled by the
spectral band width corresponding to the IC glass, for
the IC glass at $T=0$ (solid line), $\sigma$ phase at $T=0.4$ (dashed
line), and $\sigma$ phase at $T=0$ (bars).  (a) Distribution of the
traces of the diagonal 3$\times$3 blocks.  Solid circles correspond
to  matrix elements of $Z14_2$ units in the IC glass. 
(b) Distribution of the traces (with opposite sign)
of the off-diagonal 3$\times$3 blocks.}
\label{f3}
\end{figure} 

\subsection{Distribution of the matrix elements 
in the dynamical matrix}
\label{s3d}

Both level-repelling and hybridization effects have demonstrated the
strong influence of disorder on properties of disordered
and crystalline modes, thus indicating a strong degree of disorder in
the IC glass.
A possible way to quantify the degree of disorder is to compare the
distribution of the dynamical-matrix elements in the IC glass with
that in the $\sigma$ phase (see Fig.~\ref{f3}).
The positive-definite dynamical matrix consists of $D\times D$ blocks
($D$ stands for the dimensionality, $D=3$ in this case).
The elements of the diagonal blocks are subject to  sum-rule
correlations with the elements of the off-diagonal blocks
\cite{Maradudin_TLDHA}.
Therefore, there are four distinct distributions of the matrix
elements: diagonal and off-diagonal elements in diagonal and
off-diagonal blocks \cite{Taraskin_01:JPCM_CPA}.
In Figs.~\ref{f3}(a)-(b), we have plotted the representative
distributions, $\rho_{\rm diag}^{\rm diag}$ and $\rho_{\rm diag}^{\rm
off}$, for the traces of the diagonal elements in diagonal and off-diagonal blocks,
respectively. 
The trace of the off-diagonal  block 
(taken with opposite sign)  for atom $i$ 
interacting with atom $j$ can be associated 
with the effective force constant, 
$\kappa_{ij} = -\sum_\alpha D_{i\alpha,j\alpha}$,  
 for interactions  
between these atoms. 
Similarly the trace of the diagonal block for atom $i$ describes the
effective force constant, $\kappa_{ii} = \sum_\alpha
D_{i\alpha,i\alpha}$, for interactions of this atom with all other
atoms.
Two conclusions can be made from a comparison of these distributions
for the IC glass and the $\sigma$ phase.  
First, the distributions for the glass can be imagined as being generically
obtained from those for the crystal by broadening the corresponding
$\delta$-functions.
Second, this broadening is strong enough so that no traces of the
individual $\delta$-functions remain; this is an indication of a strong
degree of disorder.

A natural question is, can we actually reproduce the
distributions for the glass from the crystalline ones?
The simplest way to try to do this is to introduce 
thermal (positional) disorder in the
$\sigma$ phase by increasing the temperature, and analyzing the
dynamical matrix corresponding to the instantaneous non-equilibrium
state in which all the atoms are displaced about their crystalline equilibrium
positions (positional disorder).
We have indeed found that, 
for such a force-constant disordered case, 
 the distribution of the dynamical matrix 
elements (cf. the dashed and solid lines in Fig.~\ref{f3}(a) and (b))
and the VDOS for the thermally (positionally)-distorted $\sigma$ phase
and the IC glass \cite{Simdyankin_00} are similar.
This implies that the main features of the vibrational spectrum in a
topologically disordered system can be essentially reproduced by
introducing positional disorder in its crystalline counterpart, 
at least for the metallic-type 
system 
studied here. 
The existence of topological disorder, 
as distinguished from positional disorder 
about equilibrium crystalline positions, 
does not seem to play a major role in determining  
the character of the vibrational modes, at least in this case. 
However, we have not directly addressed the situation where 
there are differences in the topological connections of atoms with 
constant force constants. 

We have also found that the distribution of the matrix elements for
atoms belonging to $Z14_2$ units involved in the present analysis 
is similar to the total distribution (see Fig.~\ref{f3}(a)).
This confirms the representative character of the $Z{14}_2$ units for which
we have performed the local projectional analysis.

\section{Conclusions}
\label{s4}

To conclude, we have presented evidence that the vibrational modes
in a topologically disordered glass are generically related to the
eigenmodes in the corresponding crystalline counterpart via
disorder-induced level-repelling and hybridization effects.
In particular, the extra states in the low-frequency regime (boson
peak) appear to correspond to crystalline states in the vicinity of
the lowest van Hove singularities in the crystalline VDOS.
We have proposed a way of defining the degree of disorder 
(weak or strong) by
comparing the distribution of the dynamical matrix elements for a 
topologically disordered structure with  its crystalline counterpart.
It appears that the main features of the vibrational spectrum of a
topologically disordered (metallic) glass may be reproduced by the
vibrational dynamics of its positionally (thermally) distorted
crystalline counterpart.

\section*{Acknowledgments}

S.I.S., M.E. and M.D. thank Trinity College, Cambridge, U.K. for
hospitality, and acknowledge support from the following Swedish
research funds: Natural Science Research Foundation (NFR), Technical
Research Foundation (TFR), and Network for Applied Mathematics (NTM).
S.N.T. is grateful to EPSRC for support.


\begin{thebibliography}{10}
\bibitem{background}
There is a non-zero background effect (due to
the non-orthogonality of the local parts of $\u^{(l)}_{\k\beta,j}$ and
$\u^{(m)}_{\mu,j}$) in the calculation of $\omega^{(1)}_{\rm cryst}$
which is clearly seen (solid circles in Fig.~\ref{f1}) and is well
defined for $\omega_{\mu} \equiv \omega_{\rm dis} \alt 10$.  For each
value of $\omega_{\rm dis}$, the background level can be estimated as
the mean value of ${\overline A}(\omega_{\rm cryst},\omega_{\rm dis})$
for $\omega_{\rm cryst} > \omega_0$, where $\omega_0$ is an arbitrary
cut-off frequency.  The background can then be subtracted from
${\overline A}(\omega_{\rm cryst},\omega_{\rm dis})$ and the
corresponding renormalized value of first moment, $\omega^{(1)}_{\rm
cryst}$, can be calculated while assuming that ${\overline
A}(\omega_{\rm cryst}>\omega_0,\omega_{\rm dis}) = 0$. The sensitivity
of this method can be assessed by changing the value of
$\omega_0$. Fig.~\ref{f1}(b) shows that the low-$\omega_{\rm dis}$ data
are the least sensitive with respect to the definition of the
background level.
\end{thebibliography}

\end{document}